\documentclass[aps,prb,twocolumn,showpacs,groupedaddress]{revtex4-2}
\usepackage{graphicx}
\usepackage{epstopdf} 
\begin{document}

\title{Self-excitation of radio waves in the metal-insulator-metal structure\\
doped with metal nanowires}

% \affiliation command applies to all authors since the last
% \affiliation command. The \affiliation command should follow the
% other information
% \affiliation can be followed by \email, \homepage, \thanks as well.
\author{V.G.~Bordo}
\email{bordo@sdu.dk}

\affiliation{Centre for Industrial Electronics, Department of Mechanical and Electrical Engineering, University of Southern Denmark, Alsion 2, DK-6400 S{\o}nderborg, Denmark}

%Collaboration name if desired (requires use of superscriptaddress
%option in \documentclass). \noaffiliation is required (may also be
%used with the \author command).
%\collaboration can be followed by \email, \homepage, \thanks as well.
%\collaboration
%\noaffiliation

\date{\today}

\begin{abstract}
A theory of self-excitation in the metal-insulator-metal structure doped with metal nanowires is developed for the case where the power is provided by an external source of radio waves. Both the transient stage of self-excitation and the steady-state regime of self-oscillation are analyzed in a fully analytical form. The numerical estimates demonstrate that this effect can be used for diverse practical purposes, in particular, for radio frequency power harvesting. These findings extend the approach developed in nano-optics to the field of electrical engineering.
\end{abstract}

% insert suggested PACS numbers in braces on next line
% insert suggested keywords - APS authors don't need to do this
%\keywords{}

%\maketitle must follow title, authors, abstract, \pacs, and \keywords
\maketitle

% body of paper here - Use proper section commands
% References should be done using the \cite, \ref, and \label commands
\section{Introduction}
The progress in nanotechnologies has considerably modified the landscape of available materials which possess the properties not attainable in the past. Nanocomposite materials, which contain nanosized inclusions, have found numerous applications in diverse fields of science, engineering and medicine. One is able to construct such artificial substances at will in order to obtain the desirable properties.\\
This avenue requires, however, a deep understanding of the relation between the structure and composition of the material, on the one hand, and its properties, on the other hand. When turning to the optical and dielectric properties of nanocomposites, one needs to know how the microscopic (local) electric field in the material is related with the dielectric properties of the host material and nanoinclusions. This problem dates back to the Maxwell Garnett approximation in which the spherical inclusions are modeled by point dipoles and their response to the electric field is described by the polarizability of a sphere \cite{Maxwell04,Markel16}.\\
The Maxwell Garnett approach was developed, however, "provided the medium under consideration extends throughout a space of dimensions which in no direction are of an order of magnitude so small as a wavelength" \cite{Maxwell04}. If this condition is not fulfilled, one should take into account that the radiation emitted by the oscillating induced dipoles of the inclusions is reflected by the medium boundaries back to the nanocomposite volume, thus modifying the local electric field \cite{Bordo18}. The impact of this effect on the optical properties of a nanocomposite slab has been analyzed in the context of the broadband perfect absorption \cite{Bordo21} and Dicke superradiance \cite{Bordo21a}.\\
For a subwavelength slab of nanocomposite material, the inclusions are confined within a wavelength from the boundaries and therefore interact with each other through their reflected fields coherently. Using the terminology adopted in electrical engineering, each nanoinclusion undergoes a positive feedback from the reflective surfaces that can lead, under certain conditions, to its self-excitation (self-oscillation) \cite{Jenkins13}. The resulting temporal exponential growth of the nanoinclusion polarization is terminated when the saturation comes into play and the steady-state regime is established \cite{Bordo16,Bordo17}.\\
The above development concerns the field of nano-optics, where the metallic inclusions are supposed to be excited at the localized surface plasmon polariton frequency. However, similar effects one can expect in the field of electrical engineering, where the operating frequencies correspond to wavelengths much larger than the linear dimensions of the nanocomposite structures. In such a case, both the capacitance and the breakdown field in the nanocomposite capacitors are largely affected by the influence of the capacitor electrodes \cite{Bordo22,Bordo22a}.\\
Similarly, one can search for a nanosturcture design in which a self-excitation process can be realized in the radio frequency range exploited in electrical engineering. Such an approach has potentially a plenty of applications. As one of them, it is worthwhile to mention radio frequency power harvesting, which has been currently actively developing \cite{Park17}. This concept is based on the so-called rectenna approach, in which radio waves are received by an antenna and the resulting electrical signal is rectified and amplified providing a DC voltage necessary for applications. A disadvantage of this technology is a requirement that the powered device should be placed not too far from the source of radio waves. If the antenna in this design is replaced with a self-excited generator, which can be powered even by very weak radio signals, this restriction is lifted.\\
In the present paper, we develop the theory of self-excitation in the metal-insulator-metal (MIM) structure with metal nanowires (NWs) being randomly dispersed in the dielectric slab. The consideration is carried out for the case where the external source provides electromagnetic power in the radio frequency range. Both the self-excitation condition and the self-oscillation steady-state regime are analyzed in a fully analytical form.\\
The paper is organized as follows. Section \ref{sec:local} introduces the local field in the MIM structure. Sections \ref{sec:evolution} and \ref{sec:steady} describe the evolution of the NW polarization in the structure in both transient and steady-state regimes, respectively. Sections \ref{sec:estimates} and \ref{sec:conclusion} provide some numerical estimates and the summary of the paper, respectively.
\section{Local field in the MIM structure}\label{sec:local}
\begin{figure}
\includegraphics[width=\linewidth]{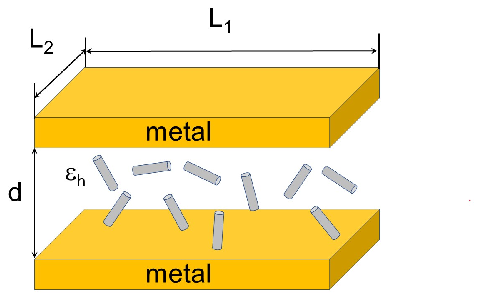}
\caption{\label{fig:structure} The sketch of the MIM structure under consideration.}
\end{figure}

Let us consider a dielectric slab of thickness $d$ and dielectric permittivity $\epsilon_h$, disposed between two metal films, in which identical metal NWs are randomly dispersed (see Fig. \ref{fig:structure}). The dimensions of both films are $L_1\times L_2$ while their thickness is unimportant, and we only assume that it is much larger than the so-called "impedance-matching thickness," which is of the order of a few nanometers and does not depend on the frequency \cite{Kaplan18}. We assume that the NWs are randomly oriented and their lengths are much less than the wavelength of interest. We choose the $z$ axis of the coordinate system along the normal to the slab with its origin at the center of the slab and the $x$ and $y$ axes along its sides of the lengths $L_1$ and $L_2$, respectively.\\
The polarizability of a NW having the dielectric permittivity $\epsilon_i$ can be obtained from the polarizability tensor of a prolate spheroid with the semi-axes $a=b$ and $c$ \cite{Landau},
\begin{equation}\label{eq:alpha}
   \alpha_{jj}=\frac{\epsilon_hV}{4\pi}\frac{\epsilon_i-\epsilon_h}{\epsilon_h+L_j(\epsilon_i-\epsilon_h)},\quad j=\xi,\eta,\zeta
\end{equation}
with the depolarization factors  
\begin{equation}
    L_{\xi}=L_{\eta}=\frac{1}{2}\left(1-L_{\zeta}\right)
\end{equation}
and
\begin{equation}
   L_{\zeta}=\frac{1-e^2}{2e^3}\left(\text{ln}\frac{1+e}{1-e}-2e\right). 
\end{equation}
Here $e=\sqrt{1-a^2/c^2}$, the coordinate axes $\xi$, $\eta$ and $\zeta$ are directed along the semi-axes $a$, $b$ and $c$, respectively, and $V=(4\pi/3)a^2c$ is the NW volume.\\ 
For a cylindrical NW $c\rightarrow\infty$ while its volume is kept finite, and $L_{\xi}=L_{\eta}=1/2$ and $L_{\zeta}=0$. Taking into account that for good conductors in a periodic field of frequency $\omega$ in the radio frequency domain $\epsilon_i=4\pi i\sigma_0/\omega$ with $\sigma_0$ being the conductivity of the metal for constant currents \cite{Landau} and $\vert\epsilon_i\vert\gg\epsilon_h$ one obtains
\begin{equation}\label{eq:alphazz}
\alpha_{\zeta\zeta}\approx \frac{1}{4\pi}\epsilon_iV=\frac{i\sigma_0}{\omega}V
\end{equation}
and
\begin{equation}
\alpha_{\xi\xi}=\alpha_{\eta\eta}\approx \frac{1}{2\pi}\epsilon_hV\ll \vert\alpha_{\zeta\zeta}\vert.
\end{equation}
We assume therefore that the NWs are uniaxial with the induced dipole moments, ${\bf p}$, directed along their axes and the polarizability given by Eq. (\ref{eq:alphazz}), i.e.
\begin{equation}\label{eq:static}
p_{\zeta}({\bf r})=\frac{i\sigma_0}{\omega}VE_{\zeta}({\bf r}),
\end{equation}
where ${\bf r}$ is the radius vector of the NW position and $E_{\zeta}$ is the effective (local) electric field component parallel to the NW axis \cite{Born70}.\\ 
The above result is valid, however, in the quasi-static limit $\omega\rightarrow 0$ for a constant field amplitude. In order to take into account the transient process which occurs during the onset of the field, one should introduce the NW dipole moment relaxation time, $\tau$, which is determined by the mean time between successive electron collisions. Then, assuming the time dependence $E_{\zeta}(t)=E_0(t)\exp(-i\omega t)$ for the local field and $p_{\zeta}(t)=p_0(t)\exp(-i\omega t)$ for the dipole moment, the equation for the dipole moment amplitude evolution takes the form \cite{Ashcroft}
\begin{equation}\label{eq:p0}
\frac{dp_0(t)}{dt}-i\omega p_0(t)+\frac{p_0(t)}{\tau}=\frac{i\sigma_0}{\omega\tau} V E_0(t).
\end{equation}
Let us note that in the steady-state regime, where $dp_0/dt=0$ and $E_0$ does not depend on time, the above equation gives $p_0=(i\sigma(\omega)/\omega) VE_0$ with $\sigma(\omega)=\sigma_0/(1-i\omega\tau)$ being the metal conductivity for an AC field in the Drude model, as it is expected from Eq. (\ref{eq:static}). However, at the transient stage one should keep $dp_0/dt$ in Eq. (\ref{eq:p0}).\\
We will be interested in the NW polarization component perpendicular to the slab, $P_z(t)=P_0(t)\exp(-i\omega t)$, which for randomly oriented NWs is found as
\begin{equation}\label{eq:P0}
P_0(t)=Np_0(t)\langle \cos^2\theta \rangle = \frac{1}{2}Np_0(t),
\end{equation}
where $N$ is the number density of NWs, $\theta$ is the angle between the NW axis and the $z$ axis and the angular brackets denote the averaging over the NW orientations. Using Eq. (\ref{eq:P0}) one can rewrite Eq. (\ref{eq:p0}) in the form
\begin{equation}\label{eq:P}
\frac{dP_0(t)}{dt}-i\omega P_0(t)+\frac{P_0(t)}{\tau}=\frac{i\sigma_0}{2\omega\tau} f E_0(t)
\end{equation}
with $f=NV$ being the volume fraction of NWs.\\
The local field in its turn satisfies the integral equation \cite{Bordo16}
\begin{equation}\label{eq:integral}
E_0({\bf r},t)=E_{ex}({\bf r})+\int_V^{\prime} F_{zz}({\bf r},{\bf r}^{\prime};\omega)P_0({\bf r}^{\prime},t)d{\bf r}^{\prime},
\end{equation}
where $E_{ex}({\bf r})\exp(-i\omega t)$ is the $z$ component of the external electric field penetrated into the nanocomposite which is assumed to be switched on at $t=0$, $V$ is the volume of the nanocomposite slab and the prime above the integral sign implies removal of the point ${\bf r}^{\prime}={\bf r}$ from the integration that isolates the singularity in Green's function (see Ref. \cite{Born70} for the detail). The kernel in this equation, $F_{zz}({\bf r},{\bf r}^{\prime})$, is the component of the so-called field susceptibility tensor which relates the electromagnetic field at the point ${\bf r}$ generated by a classical dipole with the dipole moment itself located at the point ${\bf r}^{\prime}$ \cite{Sipe84}. This quantity can be expressed in terms of Green's function of the vector wave equation in the frequency domain with appropriate boundary conditions and provides therefore a solution for inhomogeneous Maxwell's equations (see Ref. \cite{Bordo16}, Section IIIA for the detail). Equation (\ref{eq:integral}) together with Eq. (\ref{eq:P}) determines a self-consistent solution for the local field $E_0(t)$.\\
\section{Evolution of the NW polarization}\label{sec:evolution}
The field susceptibility tensor $\bar{\bf F}({\bf r},{\bf r}^{\prime};\omega)$ which enters Eq. (\ref{eq:integral}) can be split into two contributions, $\bar{\bf F}^0({\bf r},{\bf r}^{\prime};\omega)$, which originates from the dipole field \cite{Sipe79}, and $\bar{\bf F}^R({\bf r},{\bf r}^{\prime};\omega)$, which results from the dipole field reflected from the parallel metal films \cite{Nha96}. It can be represented in the form of the 2D Fourier integral as 
\begin{eqnarray}
\bar{\bf F}({\bf r},{\bf r}^{\prime};\omega)\nonumber\\
=\frac{1}{(2\pi)^2}\int\bar{\bf f}(z,z^{\prime};\omega,\kappa)e^{i\kappa_x(x-x^{\prime})}e^{i\kappa_y(y-y^{\prime})}d\kappa_xd\kappa_y\nonumber\\
\end{eqnarray}
with $\kappa=(\kappa_x^2+\kappa_y^2)^{1/2}$.\\
Taking into account that in the radio frequency domain the reflection coefficients from metal surfaces are equal with high accuracy to one, we obtain \cite{note1}
\begin{equation}\label{eq:kernel}
f_{zz}(z,z^{\prime};\omega,\kappa)=H^-(z^{\prime};\omega,\kappa)e^{-iWz}+H^+(z^{\prime};\omega,\kappa)e^{iWz}
\end{equation}
with $W=[(\omega/c)^2\epsilon_h-\kappa^2]^{1/2}$ and $c$ being the speed of light in vacuum, where
\begin{eqnarray}
H^{\pm}(z;\omega,\kappa)\nonumber\\
=\frac{i\pi\kappa^2}{\epsilon_hW}\left\{1+\frac{2e^{iWd}}{1-e^{2iWd}}\left[e^{iW(d\mp z)}+e^{\pm iWz}\right]\right\}.
\end{eqnarray}
Introducing the Fourier transform
\begin{eqnarray}
E_0({\bf r},t)\nonumber\\
=\frac{1}{(2\pi)^2}\int_{-\infty}^{\infty}\int_{-\infty}^{\infty}e(z;\kappa_x,\kappa_y;t)e^{i\kappa_xx}e^{i\kappa_yy}d\kappa_xd\kappa_y\nonumber\\
\end{eqnarray}
and the Laplace transform
\begin{equation}
\tilde{e}(z;\kappa_x,\kappa_y;s)=\int_0^{\infty}e(z;\kappa_x,\kappa_y;t)e^{-st}dt
\end{equation}
along with similar transforms for the quantities $E_{ex}$ and $P_0$, one obtains from Eqs. (\ref{eq:P}) and (\ref{eq:integral})
\begin{equation}\label{eq:ptilde}
(s-i\omega+\tau^{-1})\tilde{p}(z;\kappa_x,\kappa_y;s)=\frac{i\sigma_0}{2\omega\tau} f \tilde{e}(z;\kappa_x,\kappa_y;s)
\end{equation}
and
\begin{eqnarray}\label{eq:integral1}
\tilde{e}(z;\kappa_x,\kappa_y;s)=\frac{1}{s}e_{ex}(z;\kappa_x,\kappa_y)\nonumber\\
+\int_{-d/2}^{d/2}f_{zz}(z,z^{\prime};\omega,\kappa)\tilde{p}(z^{\prime};\kappa_x,\kappa_y;s)dz^{\prime},
\end{eqnarray}
respectively, where we have assumed that $P_0=0$ at $t=0$.\\
The degenerate form of the kernel (\ref{eq:kernel}) in the integral equation (\ref{eq:integral1}) allows one to reduce both Eqs. (\ref{eq:ptilde}) and (\ref{eq:integral1}) to a set of two linear algebraic equations, which in turn can be written in a matrix form
\begin{eqnarray}\label{eq:matrix}
\left(
\begin{array}{c}
\mathcal{E}^-\\
\mathcal{E}^+
\end{array}\right)
=\frac{1}{s}\left(
\begin{array}{c}
\mathcal{E}_{ex}^-\\
\mathcal{E}_{ex}^+
\end{array}\right)\nonumber\\
+\frac{i\eta}{s-i\omega+\tau^{-1}}\left(
\begin{array}{c}
\mathcal{H}^{--} \quad \mathcal{H}^{-+}\\
\mathcal{H}^{+-} \quad \mathcal{H}^{++}
\end{array}\right)
\left(
\begin{array}{c}
\mathcal{E}^-\\
\mathcal{E}^+
\end{array}\right)
\end{eqnarray}
with $\mathcal{E}^{\pm}$=$\int_{-d/2}^{d/2}H^{\pm}(z)\tilde{e}(z)dz$, $\mathcal{E}_{ex}^{\pm}$=$\int_{-d/2}^{d/2}H^{\pm}(z)e_{ex}(z)dz$, $\mathcal{H}^{\alpha\pm}$=$\int_{-d/2}^{d/2}H^{\alpha}(z)\exp(\pm iWz)dz$, $\alpha$=$\pm$ and $\eta$=$\sigma_0f/2\omega\tau$.\\
Equation (\ref{eq:matrix}), being transformed to the basis of the eigenvectors of the matrix 
\begin{eqnarray}\label{eq:H}
\hat{\mathcal{H}}=\left(\begin{array}{c}
\mathcal{H}^{--} \quad \mathcal{H}^{-+}\\
\mathcal{H}^{+-} \quad \mathcal{H}^{++}
\end{array}\right),
\end{eqnarray}
takes the diagonal form
\begin{eqnarray}\label{eq:diagonal}
\left(
\begin{array}{c}
\tilde{\mathcal{E}}^-\\
\tilde{\mathcal{E}}^+
\end{array}\right)
=\frac{1}{s}\left(
\begin{array}{c}
\tilde{\mathcal{E}}_{ex}^-\\
\tilde{\mathcal{E}}_{ex}^+
\end{array}\right)\nonumber\\
+\frac{i\eta}{s-i\omega+\tau^{-1}}\left(
\begin{array}{c}
\lambda_- \quad 0\\
0 \quad \lambda_+
\end{array}\right)
\left(
\begin{array}{c}
\tilde{\mathcal{E}}^-\\
\tilde{\mathcal{E}}^+
\end{array}\right),
\end{eqnarray}
where the tilde denotes the quantities in the new basis and $\lambda_{\pm}=\lambda_{\pm}^{\prime}+i\lambda_{\pm}^{\prime\prime}$ are the eigenvalues of the matrix $\hat{\mathcal{H}}$. One finds from here the solutions for the field amplitudes
\begin{equation}\label{eq:solution}
\tilde{\mathcal{E}}^{\pm}(s)=\frac{s-i\omega+\tau^{-1}}{s(s-i\omega+\tau^{-1}-i\eta\lambda_{\pm})}\tilde{\mathcal{E}}^{\pm}_{ex}.
\end{equation}
Self-excitation of the field in the MIM structure becomes possible if the time evolution of the field amplitudes, which is obtained as the inverse Laplace transform of Eq. (\ref{eq:solution}), contains exponentially increasing terms. This occurs when at least one pole of the functions $\tilde{\mathcal{E}}^{\pm}(s)$ lies in the right half-plane of the complex plane of $s$, i.e. when $\lambda^{\prime\prime}_{\pm}$ is negative and
\begin{equation}
-\eta\lambda^{\prime\prime}_{\pm}(\kappa)>\tau^{-1}
\end{equation}
or, using the definition of $\eta$,
\begin{equation}\label{eq:threshold}
-\frac{\sigma_0f}{2\omega}\lambda^{\prime\prime}_{\pm}(\kappa)>1.
\end{equation}
The inequality (\ref{eq:threshold}) provides the criterion of self-excitation which puts a lower limit (threshold) for the NW doping density $f$.
\section{Steady-state regime}\label{sec:steady}
The exponential increase in time terminates and a steady-state regime of self-oscillation is established at $t\rightarrow\infty$ when Eq. (\ref{eq:diagonal}) has a non-trivial solution in the limit $\tilde{\mathcal{E}}_{ex}^{\pm}\rightarrow 0$. The steady-state solutions for the local field amplitudes are found as
\begin{equation}\label{eq:limit}
\lim_{t\rightarrow\infty}\tilde{\mathcal{E}}^{\pm}(t)=\lim_{s\rightarrow 0}s\tilde{\mathcal{E}}^{\pm}(s)=\frac{-i\omega+\tau^{-1}}{-i\omega+\tau^{-1}-i\eta\lambda_{\pm}}\tilde{\mathcal{E}}^{\pm}_{ex}.
\end{equation}
Equation (\ref{eq:limit}) provides a non-zero solution for $\tilde{\mathcal{E}}^{\pm}$ at vanishing values of $\tilde{\mathcal{E}}^{\pm}_{ex}$ only if the denominator in it equals zero that happens if $-\eta\lambda_{\pm}^{\prime}=\omega$ and $-\eta\lambda_{\pm}^{\prime\prime}=\tau^{-1}$ simultaneously or, using the definition of $\eta$,
\begin{equation}\label{eq:steady1}
-\frac{\sigma^s_0f}{2\omega}\lambda^{\prime}_{\pm}(\kappa)=\omega\tau
\end{equation}
and
\begin{equation}\label{eq:steady2}
-\frac{\sigma^s_0f}{2\omega}\lambda^{\prime\prime}_{\pm}(\kappa)=1,
\end{equation}
where $\sigma^s_0$ denotes the NW conductivity in the steady-state regime.\\ 
The value of $\sigma^s_0$ is determined by Joule heating caused by the oscillating electric current, which, along with the electric field, is increasing during the self-excitation process \cite{Cheng15}. As far as $\sigma^s_0<\sigma_0$ the latter equation is consistent with the threshold condition (\ref{eq:threshold}). Being initially prepared in the state above the threshold, the MIM structure evolves to the steady-state determined by Eqs. (\ref{eq:steady1}) and (\ref{eq:steady2}), which dictate both the established field spatial modes, specified by $\kappa$, and NW temperature. The temperature, in its turn, determines the largest field amplitude which can be attained in the self-excitation process.\\
Dividing Eq. (\ref{eq:steady1}) by Eq. (\ref{eq:steady2}), one obtains the equation which provides for a given $\omega$ the established spatial modes of self-oscillation
\begin{equation}\label{eq:ratio}
\frac{\lambda_{\pm}^{\prime}(\kappa)}{\lambda_{\pm}^{\prime\prime}(\kappa)}=\omega\tau.
\end{equation}
The straightforward calculations give for the eigenvalues of the matrix $\hat{\mathcal{H}}$
\begin{equation}
\lambda_-(\kappa)=-\frac{\pi\kappa^2d}{\epsilon_hW}\frac{e^{iWd/2}}{\sin(Wd/2)}\left[1+\frac{\sin(Wd)}{Wd}\right]
\end{equation}
and
\begin{equation}
\lambda_+(\kappa)=-\frac{i\pi\kappa^2d}{\epsilon_hW}\frac{e^{iWd/2}}{\cos(Wd/2)}\left[1-\frac{\sin(Wd)}{Wd}\right].
\end{equation}
Let us note that for imaginary values of $W$, i.e. for $\kappa>(\omega/c)\sqrt{\epsilon_h}$, both $\lambda_-$ and $\lambda_+$ are real and the condition (\ref{eq:steady2}) cannot be fulfilled. In what follows, we assume therefore that $W$ is real, i.e. $\kappa<(\omega/c)\sqrt{\epsilon_h}$. Then Eq. (\ref{eq:ratio}) is split in two equations 
\begin{equation}\label{eq:minus}
-\cot\left(\frac{Wd}{2}\right)=\tan\left(\frac{Wd}{2}-\frac{\pi}{2}\right)=-\omega\tau
\end{equation}
and
\begin{equation}\label{eq:plus}
\tan\left(\frac{Wd}{2}\right)=-\omega\tau
\end{equation}
for the "$-$" and "$+$" branches, respectively. As far as in the radio frequency domain $\omega\tau\ll 1$, both Eqs. (\ref{eq:minus}) and (\ref{eq:plus}) are reduced to
\begin{equation}
\frac{Wd}{2}\approx\frac{\pi n}{2}
\end{equation}
with $n=1,2,3,...$, odd and even integers being corresponding to the "$-$" and "$+$" branches, respectively. Let us note that these modes are close to the waveguide modes between two perfectly conducting plates \cite{Balanis}. One finds from here that only the spatial modes which satisfy the condition
\begin{equation}\label{eq:kappa}
\kappa\approx\sqrt{\left(\frac{\omega}{c}\right)^2\epsilon_h-\left(\frac{\pi n}{d}\right)^2}
\end{equation}
can be self-oscillating. The realness of $\kappa$ limits the possible number of modes by the requirement
\begin{equation}
n<\frac{\omega d}{\pi c}\sqrt{\epsilon_h}=\frac{2d}{\lambda}\sqrt{\epsilon_h}
\end{equation}
with $\lambda$ being the wavelength of the external field in vacuum, that puts in its turn a lower limit for the slab thickness
\begin{equation}\label{eq:dmin}
d>\frac{\lambda}{2\sqrt{\epsilon_h}}\equiv d_{\min}.
\end{equation}
The steady-state value of $\sigma_0^s$ one can find from Eqs. (\ref{eq:steady2}) and (\ref{eq:kappa}). For the lowest-order mode with $n=1$ one obtains
\begin{equation}
\lambda_-^{\prime\prime}=-\frac{\pi^2}{\epsilon_h}\left[\left(\frac{d}{d_{\min}}\right)^2-1\right]
\end{equation}
and
\begin{equation}
\sigma_0^s=-\frac{2\omega}{f\lambda_-^{\prime\prime}}.
\end{equation}
For practical purposes, the value of the electric field averaged over the slab surface is of interest, i.e.
\begin{eqnarray}
\langle E_0({\bf r})\rangle=\frac{1}{L_1L_2}\int_{-L_2/2}^{L_2/2}\int_{-L_1/2}^{L_1/2}E_0({\bf r})dxdy
=\frac{1}{(2\pi)^2}\nonumber\\
\times\int_{-\infty}^{\infty}\int_{-\infty}^{\infty}e(z;\kappa)\frac{\sin(k_xL_1/2)}{k_xL_1/2}\frac{\sin(k_yL_2/2)}{k_yL_2/2}dk_xdk_y.\nonumber\\
\end{eqnarray}
The functions $\sin X/X$ in the integrand are essentially nonzero in the range $\vert X\vert\le 3$, i.e. the essential range of integration is limited to the ranges $\vert k_x\vert\le 6/L_1$ and $\vert k_y\vert\le 6/L_2$. Assuming for simplicity that $L_1=L_2\equiv L$, one concludes that only the range $\kappa\le 6\sqrt{2}/L$ is significant for the averaged field. This imposes a requirement for the linear dimensions of the MIM structure
\begin{equation}\label{eq:Lmax}
L\le\frac{6\sqrt{2}}{\kappa}=\frac{6\sqrt{2}d}{\pi\sqrt{\left(d/d_{\min}\right)^2-1}}.
\end{equation}
\section{Numerical estimates}\label{sec:estimates}
We perform some numerical estimates for a MIM structure doped with Ag NWs for which the experimental data are available. For estimates we take $\nu=\omega/2\pi=25$ GHz, $\epsilon_h=2.25$, NW length $L_{NW}=27$ $\mu$m, and NW diameter $D_{NW}=227$ nm \cite{Cheng15}. According to Eq. (\ref{eq:dmin}) the minimum value of the nanocomposite slab thickness is $d_{\min}=0.4$ cm.\\
The experimentally determined temperature dependence of the Ag NW resistivity is given by \cite{Cheng15}
\begin{eqnarray}
\rho_0(T)\equiv\sigma_0^{-1}(T)\nonumber\\
=3.25\times 10^{-8}\Omega\cdot\text{m}+1.68\times 10^{-10}\Omega\cdot\text{m}/\text{K}\times T.
\end{eqnarray}
Taking $T=293$ K and $d=0.5$ cm one finds from here $\sigma_0=1.22\times 10^7$ S/m = $1.1\times 10^{17}$ s$^{-1}$ and the threshold value of the NW doping density $f_{\min}=1.2\times 10^{-6}$. From Eq. (\ref{eq:Lmax}) one obtains the maximum dimension of the MIM structure $L_{\max}=1.8$ cm.\\
The NW temperature rise under an applied electric current $I$ can be estimated as \cite{Cheng15}
\begin{equation}
\Delta T=\frac{I^2RL_{NW}}{12kA},
\end{equation}
where $R$ is the NW resistance, $k=191.5$ W/K$\cdot$m is the NW thermal conductivity and $A=\pi D^2_{NW}/4$ is the NW cross section. Taking into account that the allowable temperature increase is limited by the Ag melting point $T_m=1235$ K and expressing $I=\sigma_0 AE$ one obtains for the attainable field amplitude $E_{\max}=160$ V/cm that for $d=0.5$ cm gives the voltage amplitude $U_{\max}=80$ V. As it follows from Eq. (\ref{eq:steady2}), this value can be achieved in the steady-state regime if $f=3.4\times 10^{-6}$.
\section{Conclusion}\label{sec:conclusion}
In this paper, we have developed the theory of self-excitation in the MIM structure doped with metal NWs which can be initiated by radio waves of vanishing amplitude. We have found the threshold condition for the NW volume fraction and the condition which determines the steady-state electric field amplitude in analytical forms. The numerical estimates carried out for Ag NWs have demonstrated that this effect can provide a significant voltage between the two metal films that can be exploited, for example, for radio frequency power harvesting.\\


\begin{thebibliography}{99}
\bibitem{Maxwell04} J.C.~Maxwell Garnett, Colours in metal glasses and in metallic films, Philos. Trans. R. Soc. London A {\bf 203}, 385 (1904).
\bibitem{Markel16} V.A.~Markel, Introduction to the Maxwell Garnett approximation: Tutorial, J. Opt. Soc. Am. A {\bf 33}, 1244 (2016).
\bibitem{Bordo18} V.G.~Bordo, Local field in finite-size metamaterials: Application to composites of dielectrics and metal nanoparticles, Phys. Rev. B {\bf 97}, 115410 (2018).
\bibitem{Bordo21} V.G.~Bordo, Theory of light reflection and transmission by a plasmonic nanocomposite slab: emergence of
broadband perfect absorption, J. Opt. Soc. Am. B {\bf 38}, 1442 (2021).
\bibitem{Bordo21a} V.G.~Bordo, Dicke superradiance from a plasmonic nanocomposite slab, J. Opt. Soc. Am. B {\bf 38}, 2104 (2021).
\bibitem{Jenkins13} A.~Jenkins, Self-oscillation, Phys. Rep. 525, 167 (2013).
\bibitem{Bordo16} V.G.~Bordo, Self-excitation of surface plasmon polaritons, Phys. Rev. B {\bf 93}, 155421 (2016).
\bibitem{Bordo17} V.G.~Bordo, Proposal for a self-excited electrically driven surface plasmon polariton generator, Appl. Phys. Lett. {\bf 110}, 033110 (2017).
\bibitem{Bordo22} V.~Bordo and T.~Ebel, How to determine the capacitance of a nanocomposite capacitor, AIP Adv. {\bf 12}, 045107 (2022).
\bibitem{Bordo22a} V.~Bordo and T.~Ebel, Theory of Electrical Breakdown in a Nanocomposite Capacitor, Appl. Sci. {\bf 12}, 5669 (2022).
\bibitem{Park17} L.-G.~Tran, H.-K.~Cha, and W.-T.~Park, RF power harvesting: a review on designing methodologies and applications, Micro and Nano Syst. Lett. {\bf 5}, 14 (2017).
\bibitem{Kaplan18} A.E.~Kaplan, Metallic nanolayers: a sub-visible wonderland of optical properties, J. Opt. Soc. Am. B {\bf 35}, 1328 (2018).
\bibitem{Landau} L.D.~Landau, E.M.~Lifshitz and L.P.~Pitaevskii {\it Electrodynamics of Continuous Media} (Elsevier, Amsterdam, 1984).
\bibitem{Born70} M.~Born and E.~Wolf, {\it Principles of Optics} (Pergamon Press, Oxford, 1970).
\bibitem{Ashcroft} See N.W.~Ashcroft and N.D.~Mermin, {\it Solid State Physics} (Harcourt, Fort Worth, 1976), where a similar equation is introduced for the current density in the Drude model.
\bibitem{Sipe84} J.M.~Wylie and J.E.~Sipe, Quantum electrodynamics near an interface, Phys. Rev. A {\bf 30}, 1185 (1984).
\bibitem{Sipe79} J.E.~Sipe, The ATR spectra of multipole surface plasmons, Surf. Sci. {\bf 84}, 75 (1979).
\bibitem{Nha96} H.~Nha and W.~Jhe, Cavity quantum electrodynamics between parallel dielectric surfaces, Phys. Rev. A {\bf 54}, 3505 (1996).
\bibitem{note1} Following Ref. \cite{Bordo21a}, we have approximated the $\Theta$ function in $\bar{\bf F}^0({\bf r},{\bf r}^{\prime};\omega)$ by $1/2$ for a thin nanocomposite slab.
\bibitem{Balanis} C.A.~Balanis, {\it Advanced Engineering Electromagnetics} (Wiley, New York, 1989).
\bibitem{Cheng15} Z.~Cheng, L.~Liu, S.~Xu, M.~Lu, and X.~Wang, Temperature Dependence of Electrical and Thermal Conduction in Single Silver Nanowire, Sci. Rep. {\bf 5}, 10718 (2015).

   

\end{thebibliography}
\end{document}